\documentclass[preprint,showpacs,preprintnumbers,amsmath,amssymb]{revtex4}
\usepackage{mathrsfs}
\usepackage{graphicx,psfrag}
\usepackage{amsmath,amsfonts,amssymb}

\begin{document}
\title{Almost-Killing conserved currents: \\
a general mass function}
\author{Milton Ruiz$^1$, Carlos Palenzuela$^2$ and Carles Bona$^1$}
\affiliation{
{$^1$ Departament de Fisica, Universitat de les Illes Balears.
Palma de Mallorca, Spain}\\ {$^{2}$Canadian Institute for
Theoretical Astrophysics. Toronto, Ontario M5S 3H8, Canada}\\}


\date{\today}


\begin{abstract}
A new class of conserved currents, describing non-gravitational energy-momentum density, is presented. The proposed currents do not require the existence of a (timelike) Killing vector, and are not restricted to spherically symmetric spacetimes (or similar ones, in which the Kodama vector can be defined). They are based instead on almost-Killing vectors, which could in principle be defined on generic spacetimes.
We provide local arguments, based on energy density profiles in highly simplified (stationary, rigidly-rotating) star models, which confirm the physical interest of these {\it almost-Killing currents}. A mass function is defined in this way for the spherical case, qualitatively different from the Hern\'andez-Misner mass function. An elliptic equation determining the new mass function is derived for the Tolman-Bondi spherically symmetric dust metrics, including a simple solution for the Oppenheimer-Schneider collapse. The equations for the non-symmetric case are shown to be of a mixed elliptic-hyperbolic nature.
\end{abstract}




\maketitle


\section{Introduction}

In a previous paper~\cite{Ruiz:2012te}, we studied the role of the ergosphere in the Blandford-Znajeck mechanism. The essential tool for to identify jet formation in our numerical simulations was the energy flux density of the electromagnetic field. The spacetime geometry was given by a family of stationary and axisymmetric star models~\cite{Ansorg01,Ansorg:2003br}. This allowed us to use the timelike Killing vector for constructing the energy-momentum conserved current of the electromagnetic field. The energy flux density could then be identified in a physically sound way. Extending this approach to dynamical (non stationary) cases is a challenge, because of the lack of a well defined conserved current for generic spacetimes.

Energy conservation is one of the most outstanding physical paradigms. In General Relativity, it can be formulated through the covariant equation
\begin{equation}\label{Tab_conserv}
  \nabla_b T^{ab} = 0\,,
\end{equation}
where the stress-energy tensor contains all forms of energy of non-gravitational origin: it vanishes in vacuum, even if gravitational waves can propagate there. Gravitational energy is not included because in General Relativity gravitation is rather described by the curvature of spacetime. Nevertheless, the gravitational field gets coupled to the matter fields and this coupling does not allow to write down (\ref{Tab_conserv}) as an integral conservation law. $T^{ab}$ is indeed a two-tensor and the vanishing of its covariant divergence contains source terms in curved spacetimes. Conserved quantities, such as energy and momentum, cannot be properly defined in the standard way (by means of the divergence theorem) because the source terms turn the required integral conservation law into a balance law, with a bulk contribution depending on the (non-covariant) connection coefficients.

A conserved current can be obtained, however, when the spacetime admits a continuous symmetry, associated to some Killing vector field (KV) $\zeta$. In this case, the vector current
\begin{equation}
J^{a}={T^a}_b\,\zeta^b\,,
\label{Killing_current}
\end{equation}
is conserved, allowing for the Killing equation, that is $\,\nabla_a J^a =0$.
The divergence theorem allows to obtain an integral conservation law. Note that the gravitational field is still coupled to the matter fields, but this coupling is encoded here through the Killing vector $\zeta$. When it is timelike, we can interpret the current (\ref{Killing_current}) as the energy-momentum density current associated to a freely falling observer which is momentarily at rest with respect the stationary fleet of observers. Let us stress that we are still dealing with non-gravitational energy, which we will call for short {\it mass} in what follows, as the current (\ref{Killing_current}) vanishes in vacuum.

Note also that the current (\ref{Killing_current}) is not the only one which can be useful in the KV case: the related current
\begin{equation}
{R^a}_b\,\zeta^b
\label{diff_current}
\end{equation}
is also conserved. Modulo a global factor, which could always be used for rescaling any KV, this alternative conserved current differs from the usual choice (\ref{Killing_current}) by a term proportional to the KV itself, namely 
\begin{equation}\label{delta J}
\triangle J^a = T/2~\zeta^a\,.
\end{equation}
In stationary spacetimes, where $\zeta=\partial_t\,$ can be interpreted as representing the co-moving fleet of observers, the difference between (\ref{Killing_current}) and (\ref{diff_current}) can be understood just as an energy density redefinition, not affecting the energy flux. The results obtained in~\cite{Ruiz:2012te} with the current (\ref{Killing_current}) would not change if we had used (\ref{diff_current}) instead.

The (non-gravitational) energy density $\mathcal{E}$ and the energy flux density $S^i$ can be identified by considering a 3+1 decomposition of the spacetime such that the metric can be  written as
\begin{equation}
ds^2=-\alpha^2~dt^2 + \gamma_{ij}~(dx^i+\beta^i~dt)\,
(dx^j+\beta^j~dt)\,,
\label{eq:3+1metric}
\end{equation}
so we get
\begin{equation}
  \sqrt{\gamma}\,\mathcal{E} = -\sqrt{-g}\,J^0 \,,
  \qquad \sqrt{\gamma}\,\mathcal{S}^i = -\sqrt{-g}\,J^i \,,
\end{equation}
where $\sqrt{\gamma}$ is the space volume element. In adapted coordinates, where the timelike KV of  stationary spacetimes can be written as $\zeta^a = \delta^a_{(t)}$, the differential current
(\ref{delta J}) would contribute to the resulting energy density but not to the  energy-flux density.
Note that, although it could be useful in some cases to include the geometrical factor $\sqrt{\gamma}$ into the Energy density definition, we rather prefer to work here with a quantity which is independent of the space coordinates system, namely
\begin{equation}
\mathcal{E}= -n_aJ^a = -\alpha J^0\,,
\label{Energy_density}
\end{equation}
where $n_a=\alpha\,\delta_a^{(t)}$ stands for the future pointing timelike unit normal (up to a sign, it can be interpreted as the four-velocity of a non-rotating fiducial observer).

The extension of the conserved current (\ref{Killing_current}) to dynamical spacetimes can be done in some special cases. In the spherically symmetric case, the warped structure of the spacetime allows the use of the Kodama vector $K$ as a replacement for the missing KV~\cite{Kodama:1979vn}. The resulting {\it Kodama current}
\begin{equation}
J_{Kodama}^{a}={T^a}_b\,K^b\,,
\label{Kodama_current}
\end{equation}
is still conserved and leads to a definition of mass which allows to recover the Hern\'andez-Misner mass function~\cite{Hernandez66,Abreu:2010ru} (see section~\ref{sec:spherical case} or details). In the vacuum (Schwarzschild) case, the Kodama vector $K$ coincides with the standard KV. Also, in dust-filled spacetimes, the {\it matter current}
\begin{equation}\label{dust}
 J_{dust}^{a}={T^a}_b\,u^b\,,
\end{equation}
is conserved, where $u$ is the four-velocity of the particles.

Unfortunately, these ideas do not work for generic dynamical scenarios, where one has to deal with different types of matter and fields in non-symmetric spacetimes. In these cases, the idea of approximate symmetry, or the almost-Killing vector fields (AKV), could be a starting point to build up conserved currents, which could have then some physical interest. In this paper we propose to consider the {\it almost-Killing current}
\begin{equation}\label{AK current}
  J^a_{AK}=\frac{1}{4\pi}\,{R^a}_b\,\xi^b\,,
\end{equation}
which is conserved if $\xi$ is an almost-Killing vector field, as we will show in the next section (the $1/4\pi\,$ factor is introduced for further convenience). We provide local arguments, based on the energy density profiles of some simple star models, supporting the choice of the almost-Killing current (\ref{AK current}), rather than the standard choice (\ref{Killing_current}), for describing mass conservation in many physical scenarios.

Spherically symmetric metrics are considered in section~\ref{sec:spherical case},
either in the static case (the Schwarzschild constant density star) or in the
dynamical one (Tolman-Bondi dust solutions). A mass function
can be defined due to the essentially 1D character of the problem. In the static
case, we show that it coincides with the standard Hern\'andez-Misner mass
function. This is no longer true in the dynamical (dust) case, where we
provide a single elliptic-type equation determining the mass-energy function obtained
from the AK current $J_{AK}$. The space-homogeneous (Friedmann-Robertson-Walker)
case is considered in the Oppenheimer-Snyder collapse scenario. A solution is
obtained for the mass function which is qualitatively different from the
Hern\'andez-Misner mass. This function may be of  interest in studying
local density perturbations in a cosmological background.

Non-spherical spacetimes are finally considered in section~\ref{sec:genr spc}.
In the stationary case, more specifically for rigidly-rotating axially-symmetric star models, our results confirm the advantage  of taking the almost-Killing current $J^a_{AK}$
as the starting point for physical applications. In the generic, non-symmetric
case, we write down the full set of equations, which has a mixed
elliptic-hyperbolic character. We discuss all these results in the final section.


\section{Almost-Killing vector fields}
\label{sec:AKV}
Spacetimes with continuous symmetries can be characterized by
the fact that they admit non-trivial solutions to the Killing equation
(KE), namely
\begin{equation}
\mathcal{L}_\zeta \,g_{ab} =
\nabla_a\zeta_b + \nabla_b\zeta_a = 0\,.
\label{Killing}
\end{equation}
Since most astrophysical scenarios do not satisfy the above requirement,
there has been a considerable amount of work on constructing intrinsic
ways of characterizing {\it almost-symmetric} spacetimes. A precise
implementation of the concept of almost-symmetry has been provided by
Matzner~\cite{Matzner:1968}. Starting from a variational principle, it
defines a {\it measure} of the symmetry deviation of any given spacetime.
This idea has been used by Isaacson to study high frequency gravitational
waves in which, by defining a steady coordinate system, the radiation
effects can be easily separated from the background metric
\cite{PhysRev.166.1263}. Zalaletdinov related this symmetry deviation with
a measure of the inhomogeneity of the spacetime which can be related with
some entropy concept~\cite{Zalaletdinov99}. More recently Bona {\it et. al.}
have shown that harmonic almost-Killing motions provide a convenient
generalization of the standard harmonic motions~\cite{PhysRevD.72.124010}.

Vector fields verifying the almost-Killing equation (AKE)
\begin{align}
\nabla_b\left[\nabla^a\xi^b +\nabla^b\xi^a-
{\mu}\,(\nabla\cdot\xi)\,g^{ab}\right]=0\,,
\label{AKE}
\end{align}
with  $\mu$ a given  constant, provide an interesting generalization
of the KV~\cite{Taubes78,PhysRevD.72.124010}. Of course, any solution of
KE is also a solution of the AKE. 
Moreover, as pointed out by York~\cite{York74}, any vector field that
asymptotically satisfies the KE is, in general, asymptotically a solution
of Equation~(\ref{AKE}). The AKE can be obtained by minimizing, on a
fixed background, the Lagrangian density~\cite{PhysRevD.72.124010}
\begin{align}
L=\nabla^{(a}\xi^{b)}\,\nabla_{(a}\xi_{b)}-\frac{\mu}{2}\,
(\nabla\cdot\xi)^2\,,
\label{Lagrange_AKV}
\end{align}
which can be interpreted as a generic invariant measure of the deviation
from the strict Killing symmetry condition. The parameter $\mu$ measures the
relative weight of the two quadratic scalars in the Lagrangian.

Notice that, by commuting the  covariant derivatives in (\ref{AKE}),
the AKE can also be expressed as a generalized wave equation, namely
\begin{align}
\square\,\xi^a + {R^a}_b\,\xi^b
+ (1-\mu)\,\nabla^a(\nabla\cdot\xi)= 0\,.
\label{AKV2}
\end{align}
It follows that, for any given spacetime, the  initial value problem of the
above  equation is a standard Cauchy problem in the generic case $\mu\neq 2$
(i.e. a second order partial-differential-equation system for the four
vector components).  In particular, if $\mu=1$, the principal part of the AKE
is harmonic which, in adapted coordinates,
implies~\cite{PhysRevD.72.124010}
\begin{align}
g^{bc}\,\partial_t\,{\Gamma^a}_{bc} = 0\,.
\end{align}
In asymptotically flat spacetimes, this condition asymptotically coincides with
the so-called {\it minimal distortion shift condition} used in numerical relativity
to minimize changes in the shape of volume elements of the spacetime during
evolution~\cite{PhysRevD.17.2529}.

\subsection*{Almost-Killing mass function}

Allowing for (\ref{AKV2}), the AKE can be also interpreted as providing an explicit expression for the {\it almost-Killing current} $J_{AK}$, associated with the AKV field $\xi$,
namely
\begin{equation}\label{AKV3}
4\pi\,J^a_{AK} \equiv {R^a}_b\,\xi^b = \nabla_b\nabla^{[a}\xi^{b]} - (1-{\mu}/2)~\nabla^a(\nabla\cdot\xi)\,.
\end{equation}
This current is identically conserved if and only if the divergence of the AKV verifies
\begin{equation}\label{div constant}
\Box\, (\nabla\cdot \xi) = 0\,,
\end{equation}
as it its the case for both KV and homothetic vectors, which are just particular AKE solutions.

If this is not the case, only quasilocal quantities could be obtained from (\ref{AKV3}), unless we take the parameter choice $\mu=2$ (notice that the value of the parameter $\mu$ is irrelevant if $\xi$ is actually a KV or an homothety). Therefore, in what follows we will take
\begin{equation}
J^a_{AK} \equiv \frac{1}{4\pi}~{R^a}_b\,\xi^b = \frac{1}{4\pi}~\nabla_b\nabla^{[a}\xi^{b]}
\label{AK current detail}
\end{equation}
unless otherwise stated.

The current defined by the right-hand side of (\ref{AK current detail}) is actually the Komar current associated to the vector field~$\xi$~\cite{PhysRev.113.934,Wald84} and, by construction, it is divergence-free. The total mass $M$ contained in a closed  space volume $V$ on a constant time slice $\Sigma$ is given by
\begin{equation}
M = \int_\Sigma\,\mathcal{E}~dV = -\int_\Sigma \alpha \sqrt{\gamma}\,J^0 d^3x\,.
\label{total_energy}
\end{equation}
Allowing for (\ref{AK current detail}), the above expression  can be written as a surface integral by means of the divergence theorem, namely
\begin{equation}
M_{AK} = \frac{1}{4\pi}\int_{\partial\Sigma}\,n_b\,\nabla^{[a}\xi^{b]}~d\sigma_a\,,
\label{total_energy_detail}
\end{equation}
where $d\sigma_b$ is the oriented area element for the spacelike surface limiting the volume $V$.
We have seen how the almost-Killing approach allows to define conserved currents and
quasi-local quantities such as $M_{AK}$ in Eq.~(\ref{total_energy_detail}). However, there is an issue of choice. There are infinitely many AK currents, one for every choice of the seed vector field $\xi$. The resulting conserved current represents the energy-momentum density associated to the fleet of observers with worldlines aligned with the selected AKV vector field. A judicious choice of the seed AKV is then required in order to get a physically sound interpretation. If the spacetime admits either a timelike KV (or homothetic vector), the physical meaning is clear, but additional criteria must be considered otherwise.


\section{Spherically Symmetric Spacetimes}
\label{sec:spherical case}
The line element of a spherically symmetric spacetime can always be written as
\begin{align}
ds^2= -A(r,t)^2\,dt^2+ B(r,t)^2\,dr^2+Y(r,t)^2\,
d\Omega^2\,,
\label{metricSpherical}
\end{align}
which shows a warped-product structure. If we restrict ourselves to
transformations preserving spherical symmetry, the area radius $Y(r,t)$
becomes an invariant scalar. Therefore, it is possible to obtain from its gradient a second invariant scalar, given by
\begin{equation}
g^{ab}\,\partial_aY\,\partial_bY \equiv 1- \frac{2m}{Y}\,,
\label{HernandezMass}
\end{equation}
which can be considered as a definition of the Hern\'andez-Misner mass
function $m(r,t)$. In Schwarzschild coordinates, where the radial coordinate is chosen such that $Y(r,t)=r$, then
\begin{align}
ds^2= -A(r,t)^2\,dt^2+ \frac{dr^2}{1-2\,m(r,t)/r}
+r^2\,d\Omega^2\,.
\label{metricSchwarzschild}
\end{align}
In the stationary case, scalar invariants must be preserved. It means
that  the associated KV $\zeta$ must be orthogonal to the gradient of the area
radius $Y$ and therefore, in Schwarzschild coordinates, one can take $\zeta = \partial_t$.

On the other hand, in the non-stationary case, it is possible to define the Kodama
vector $K$ in an invariant way~\cite{Kodama:1979vn}: it must be orthogonal to the gradient of $Y$,
and normalized such that its squared norm coincides with (\ref{HernandezMass}). 
A simple calculation shows that, in Schwarzschild
coordinates, the Kodama vector is given by
\begin{equation}
K = \frac{\sqrt{1-2\,m(r,t)/r}}{A(r,t)}~\partial_t\,.
\label{Kodama}
\end{equation}
The associated conserved current can
be expressed in Schwarzschild coordinates as
\begin{equation}
J^a_ {Kodama} = T^a_{~b}\,K^b = \frac{\sqrt{1-2\,m(r,t)/r}}
{A(r,t)}~T^a_{~b}\,\zeta^b\,,
\end{equation}
which can be related with the Hern\'andez-Misner mass function through the
Einstein field equations as
\begin{equation}
J^a_{Kodama} = \frac{\sqrt{1-2\,m(r,t)/r}}{4\,\pi r^2\,A(r,t)}\,
(-m',\dot{m},0,0)
\label{Kodama_current_Schw}
\end{equation}
(here, and in what follows, we use a prime to indicate a radial derivative
while the time derivative is denoted by an upper dot).
According  to (\ref{Energy_density}), the associated energy densities are therefore given by
\begin{equation}
\mathcal{E}_{T}=\frac{m'\,A(r,t)}{4\,\pi\,r^2}\,,\qquad
\mathcal{E}_{Kodama} = \frac{m'\,\sqrt{1-2m(r,t)/r}}{4\,\pi\,r^2}\,,
\label{E_Kodama}
\end{equation}
where $\mathcal{E}_{T}$ stands for the energy density associated to the standard current (\ref{Killing_current}).
It turns out that, by integrating the above densities over a sphere
of area radius $r$, the resulting mass functions we obtain are, respectively
\begin{equation}
M_{T} = \int_\Sigma\frac{m'\,A(r)}{\sqrt{1-2m(r,t)/r}}\,dr\,,
\qquad M_{Kodama} = m(r,t)\,.
\label{Mass_Kodama}
\end{equation}
We then recover, in the generic case, the  well known Hern\'andez-Misner mass
for the Kodama definition, but not for the standard one, derived from the KV.

We can repeat the previous calculation for a generic almost-Killing vector $\xi$.
If we consider adapted coordinates (non-Schwarzschild ones in the generic
case) where the AKV is $\xi = \partial_t$ then it is straightforward to
show that
\begin{equation}
\mathcal{E}_{AK} = \frac{1}{4\,\pi\,B\,Y^2}\,
\partial_r\,\left(\frac{Y^2}{B}\,A'\right)\,.
\label{E_AK}
\end{equation}
The corresponding mass function is then given by
\begin{equation}
M_{AK} = \frac{Y^2}{B}\,A'\,,
\label{Mass_AK}
\end{equation}
which could be also obtained directly from (\ref{total_energy_detail}).
The physical meaning of this expression follows from a straightforward calculation of the acceleration of the AKV observers in their rest frame (the opposite of the gravitational pull)
\begin{equation}
\dot{u}^a = A~\frac{M_{AK}}{Y^2}~\hat{r}^a\,,
\label{Mass Kepler}
\end{equation}
where $\hat{r}$ stands for the unit normal in the radial direction.
As the mass function $M_{AK}$ amounts to the mass contained inside the spherical surface of area radius $Y$, we can see that the gravitational pull (the opposite of $\dot{u}$) coincides, up to the Lorentz factor $A$, with the  Newtonian expression derived from Kepler's law. This is an important result that strongly supports the use of almost-Killing currents in physical applications.


\subsection{The Schwarzschild constant-density interior solution}

Perhaps the simplest  geometry associated with a matter distribution
is the Schwarzschild star which corresponds to the interior solution
for a relativistic star with constant density. This solution is the
simplest analytic interior model for a relativistic  star. The assumption $\rho= const$,
which would correspond to a ultra-stiff equation of state,
corresponds to an incompressible fluid, with an infinite sound speed.

The metric of a static spherically  symmetric star with a constant
matter distribution can be written in the form
(\ref{metricSchwarzschild}), where the metric coefficients $A=A(r)$
and $m(r)$ are only functions of the area radius $r$, that is
\begin{equation}
A(r)=\frac{\rho }{\rho+3\,p}~\sqrt{1-2\,m(r)/r}\,,
\end{equation}
where $p=p(r)$ is the pressure profile, which is given by
\begin{align}
p(r)=\frac{\rho}{2\,A}~\left(\sqrt{1-2\,M_s\,r^2/R^{3}}
-\sqrt{1-2\,M_s/R}\right)\,,
\label{pressure_schwarschild}
\end{align}
with $R$ and $M_s$ the radius and total mass of the star, whereas
the mass function $m(r)$ is given by
\begin{equation}
m(r)=\left\{
\begin{array}{c}
\frac{4}{3}\pi\,\rho\,r^3\,,\qquad\qquad\textrm{for~}\,
r\leq R\,,\\
M_{s}=\frac{4}{3}\pi\,\rho\,R^3\,,\qquad\textrm{for~}\,
r\geq R\,.
\label{m_schwarschild}
\end{array}\right.
\end{equation}
Since  the spacetime is both spherical and static, one can  define
conserved currents associated to either the Kodama or the Killing
vector as in the last section. So, the corresponding energy densities
are in these cases
\begin{equation}
\mathcal{E}_{Kodama} = A\,(\rho+3\,p)\,,\qquad
\mathcal{E}_{T}= A\,\rho\,.
\label{E_Kodama_Killing}
\end{equation}
Notice that the positivity of $\,\mathcal{E}_{Kodama}\,$ and $\,\mathcal{E}_{T}\,$  imply
the strong and the weak energy conditions, respectively.
The mass contained  over a sphere of area radius $r$ is then given by (\ref{Mass_Kodama}), namely.
\begin{equation}\label{M_kod}
M_{Kodama}= m(r)\,,\qquad
M_{T}=4\,\pi\,\int_0^R \frac{\rho\,r^2\,A}{\sqrt{1-2\,m(r)/r}}\,dr\,.
\end{equation}

Alternatively, we can use the KV field $~\zeta=\partial_t~$ to compute
the conserved AK current and the corresponding energy density. Thus,
according to~(\ref{Energy_density}), we obtain
\begin{align}
\mathcal{E}_{AK} = -\frac{1}{4\pi}\, {R^a}_b\,n_a\,\zeta^b
= A\,\left(\rho+3\,p\right)\,,
\label{Schwarszchild_star_den}
\end{align}
which coincides in this particular case with the Kodama result~\ref{M_kod}.
Note that the mass values obtained from these two definitions are also
different. Only through the Kodama/AK choice is possible to recover
the Hern\'andez-Misner mass function.

\begin{figure}[h]
\begin{center}
\includegraphics[width=0.6\textwidth,angle=0]{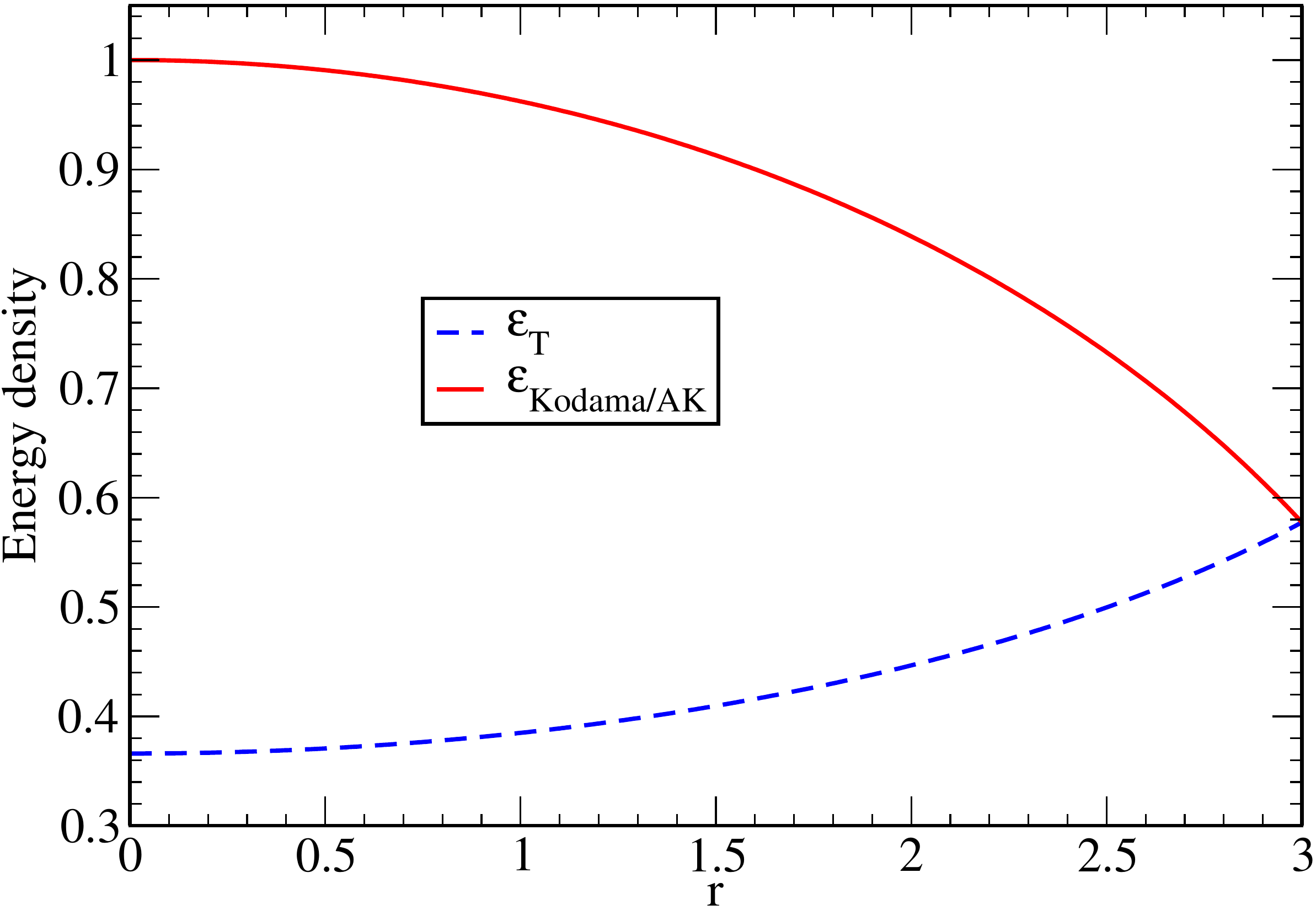}
\caption{Energy densities for a Schwarzschild star. The energy distribution computed from either the Kodama or the almost-Killing current (continuous line), has a maximum at the
center of the star that monotonically decreases until the star surfacer.  In contrast, the  energy density $\mathcal{E}_{T}$ (dashed line)
is an increasing function that reaches the maximum value at that surface.}
\label{fig:Energy_Dens}
\end{center}
\end{figure}

These energy densities are compared in Fig.~\ref{fig:Energy_Dens} for a Schwarzschild star with compactness $M/R=1/3$.
The energy density $\mathcal{E}_{AK}$ (continuous line), computed from either the Kodama or the almost-Killing currents, leads to a more natural energy distribution,
with a maximum at the star center and monotonically decreasing as one
approaches to the surface of the star. In contrast, the energy density
$\mathcal{E}_{T}$ has  the opposite and counter-intuitive behavior which
corresponds to an increasing function that reaches the maximum value at that
surface (dashed line). The geometric factor coming from the space volume element is not included in the plots.


\subsection{Tolman-Bondi dust metrics}
 The line element of a generic (spherically symmetric) dust-filled spacetime
in comoving coordinates can be written  as~\cite{Bon52,KraHer80}
\begin{equation}
ds^2= -dt^2+ \frac{Y'(r,t)^2}{1-k\,f(r)^2}~dr^2+
Y^2(r,t)~d\Omega^2\,.
\label{dust metric}
\end{equation}
The area radius $Y(r,t)$ verifies, according
to~(\ref{HernandezMass}),  the Friedmann-like equation
\begin{equation}
\dot{Y}^2 - \frac{2\,m(r)}{Y} = -k\,f(r)^2\,,
\label{Friedmann_equation}
\end{equation}
where both the Hern\'andez-Misner mass function $m(r)$
and $f(r)$ are arbitrary functions,
although the choice of $f(r)$ is restricted by the regularity requirements
of the metric. It follows from the field equations that the density $\rho$ is given by
\begin{equation}\label{dust density}
 4\,\pi\,\rho = \frac{m'}{Y^2\,Y'}\,.
\end{equation}
The vacuum case (Schwarzschild spacetime) is then recovered
where $m'=0$.

For any dust-filled spacetime, the conservation of the stress-energy tensor amounts to the conservation of the {\it matter current}
\begin{equation}\label{matter current}
  J^a_{dust} = \rho\, u^a\,
\end{equation}
where $u^a$ stands for the fluid four-velocity. The corresponding mass density is actually the matter density $\rho$. Note that the resulting mass function $M_{dust}$ is time independent (comoving coordinates) and verifies indeed
\begin{equation}\label{mass dust}
  M'_{dust} = \frac{m'}{\sqrt{1-k\,f(r)^2}}\,,
\end{equation}
which differs from the Hern\'andez-Misner mass function unless $k=0$.

Even if there is no timelike KV in the generic case, we can define other conserved currents via either the Kodama vector or the AKV. On one hand, in comoving coordinates, the
Kodama vector can be brought into the form
\begin{equation}
K^a = \sqrt{1-k\,f(r)^2}\,
(\,1\,,-{\dot{Y}}/{Y'}\,,0\,,0)\,.
\label{Kodama_FRW}
\end{equation}
The conserved Kodama current is then
\begin{equation}
J^a_{Kodama} = \sqrt{1-k\,f(r)^2}\,J^a_{dust}\,,
\end{equation}
so that the Kodama energy density can be expressed as
\begin{equation}
{\mathcal E}_{Kodama} = \frac{\sqrt{1-k\,f(r)^2}}{4\pi\,Y^2\,Y'}\,m'(r)\,,
\end{equation}
which allows to recover the Hern\'andez-Misner mass function $m(r)$ by
integrating over a sphere, as we have seen before (\ref{Mass_Kodama}).

On the other hand, according to~(\ref{AK current detail}), the conserved AK-current is given by
\begin{equation}
J^a_{AK} = \frac{\sqrt{1-k\,f(r)^2}}{4\pi\,Y^2\,Y'}\,
(-M',\dot{M},0,0)\,.
\label{AK_current}
\end{equation}
where we have introduced the potential $M(r,t)$ which coincides, up to a constant, with
the mass function (\ref{total_energy_detail}) associated to the AKV $\xi$, namely
\begin{equation}
M_{AK}= \frac{Y^2\,Y'}{\sqrt{1-k\,f(r)^2}}\,
\nabla^{[r}\xi^{t]}\,.
\label{AK_energy}
\end{equation}
The energy density associated with the AKV can be then expressed as
\begin{equation}
{\mathcal E}_{AK} = \frac{\sqrt{1-k\,f(r)^2}}{4\,\pi\,Y^2\,Y'}
\,M'\,.
\label{energy_OS}
\end{equation}

Note that, allowing for (\ref{AKV3}), the AK current can also be expressed as
\begin{equation}
J^a_{AK} = \frac{1}{4\,\pi}\,R^a_b\,\xi^b =
\rho~(\,\xi^a+2\, u\cdot\xi~u^a)\,.
\label{ricci current}
\end{equation}
In the non-vacuum case, the above expression can be inverted in
order to obtain the components of the  AKV in terms of the potential
~$M$, which are given by
\begin{equation}
\xi^a = \frac{\sqrt{1-k\,f(r)^2}}{m'}\,(M',\dot{M},0,0)\,.
\label{AKV_FRW}
\end{equation}
Plugging this result back into Eq.~(\ref{AK_energy}), we obtain a second
order elliptic PDE for the potential $M(r,t)$ that ensures that $\xi$ is a true AKV in the
non-vacuum case, that is
\begin{equation}
M-M_s  = \frac{Y^2}{2\,Y'}\,\left[~
\partial_t\, \left(~\frac{Y'^{\,2}\,\dot{M}}{m'}\right)
 + \sqrt{1-k\,f(r)^2}~\partial_r\,\left( ~\frac{\sqrt{1-k\,f(r)^2}}{m'}
\,M'\,\right)~\right]\,,
\label{mass_eq}
\end{equation}
where $M_s$ is an integration constant. Notice that the expression (\ref{AKV_FRW}) is undetermined in the vacuum case. For the Schwarzschild metric, however, a Killing vector exists, which actually coincides with the Kodama vector (\ref{Kodama_FRW}).

The time dependence in $M(r,t)$ is essential. The ansatz $\dot{M}=0$ is not compatible with the equation (\ref{mass_eq}) in the generic case. Then, it follows from (\ref{AKV_FRW}) that the AKV fleet of observers is tilted with respect to the comoving (geodesic) observers. The static value of both the 'number-of-particles' mass (\ref{mass dust}) and the Hern\'andez-Misner mass function $m(r)$ indicates that these are concepts associated to the comoving observers. From the point of view of a quasi-stationary observer, however, particles in a collapsing ball of dust are gaining kinetic energy, meaning that their mass must increase. Let us illustrate this point with a simple example


\subsubsection*{Oppenheimer-Schneider collapse}

Let us consider now the case in which the matter density distribution
$\rho$ is homogeneous. This can be interpreted as a cosmological solution,
the pressureless case of the Friedmann-Robertson-Walker (FRW) metrics such
that
\begin{equation}
Y = R(t)\,f(r)\,,\qquad m(r)=\frac{2}{9}\,f(r)^3\,.
\label{FRW}
\end{equation}
If one considers  the metric~(\ref{dust metric}) with the parameter
$k\neq -1$, the resulting spacetime corresponds to the collapse of a
homogeneous ball of incoherent matter, with a vacuum exterior metric
(Oppenheimer-Schneider collapse). On the other hand, the  value $k=+1$
allows to choose initial data for the collapse which correspond  a
momentarily static configuration. We will rather consider here for
simplicity the $k=0$ case, such that $R(t)=(t_0-t)^{2/3}$, where $t_0$ is
the time label for the collapse to the final singularity. This allows
to obtain an explicit solution of (\ref{mass_eq})
which can be written as
\begin{equation}
M(r,t) = M_s + \lambda(r)~ (t-t_0)\,,
\label{mass_FRW}
\end{equation}
where the function $\lambda(r)$ is linear in $m(r)$, as it follows from (\ref{mass_eq}).

The FRW dust interior can be matched to the vacuum exterior metric
(Schwarzschild spacetime) at the radius of the star $r=r_s$. We can adjust the linear relation between $\lambda$ and $m(r)$ in order to match the AK mass function
(\ref{mass_FRW}) with the Hern\'andez-Misner mass at the surface, namely
\begin{equation}
M(r,t)- M_s = (1-t/t_0)~\left[ \,m(r)-M_s\, \right] \,.
\label{mass_OS}
\end{equation}
This implies $\dot{M}\mid_{r_s}= 0\,$. Therefore, according with Eq.~(\ref{AK_current}), there is no flux of momentum at the surface. This means that the resulting AKV (\ref{AKV_FRW}) is comoving at the surface.

\begin{figure}[h]
\begin{center}
\includegraphics[width=0.6\textwidth,angle=0]{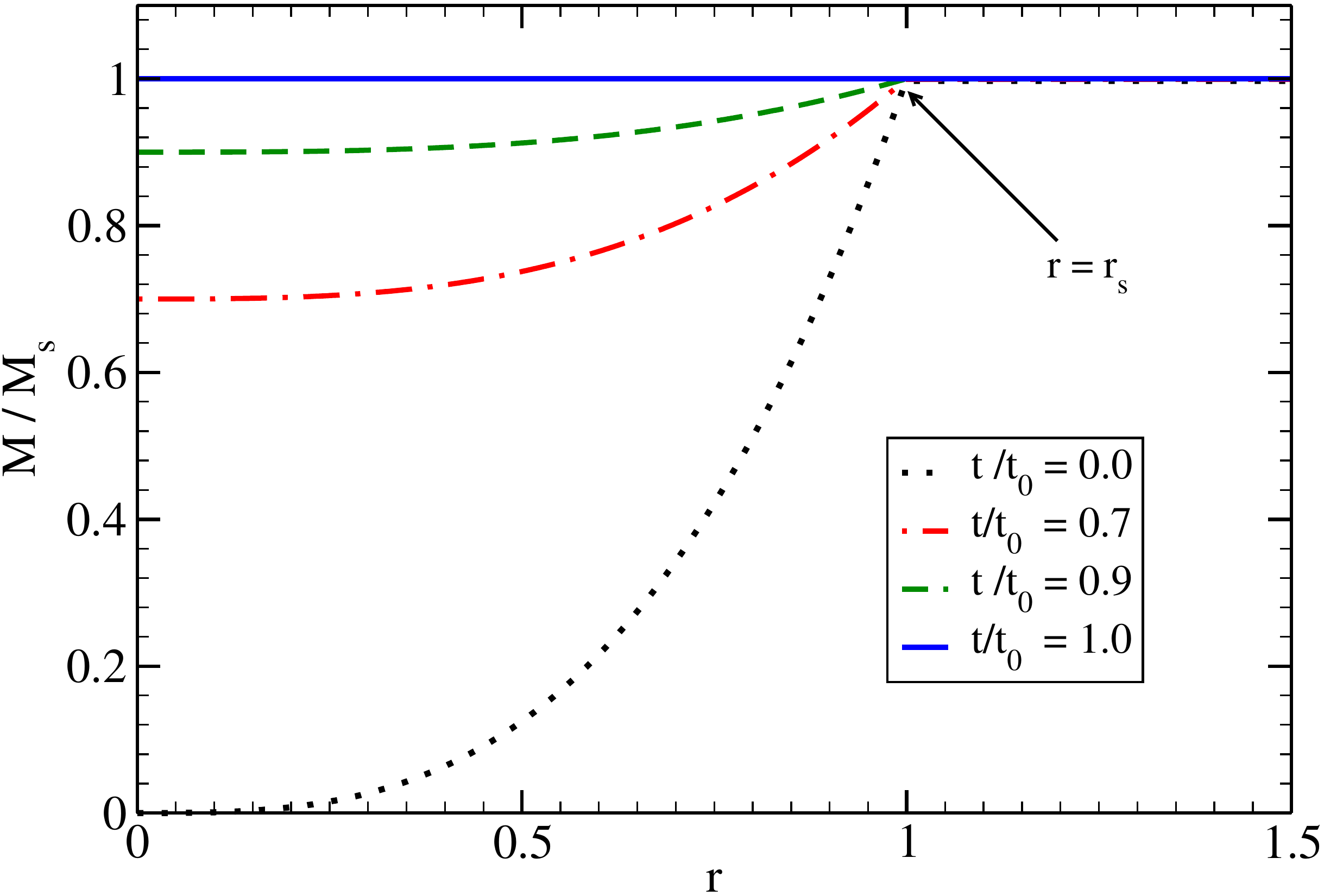}
\caption{Time evolution of the mass function $M(r,t)$
normalized to the constant $M_s$ for a FRW collapsing homogeneous ball of dust
in the flat case $k=0$. The surface is located $r_s=1.0$,
where a matching  with the Schwarzschild spacetime is enforced.
The initial mass distribution evolves towards a constant value.}
\label{fig:snapshots_Energy}
\end{center}
\end{figure}

Figure~(\ref{fig:snapshots_Energy}) displays the evolution of the mass function $M(r,t)$, normalized to the constant $M_s$, which corresponds to the mass of the Schwarzschild spacetime (we have taken $f(r)=r$ for simplicity). We can see the expected growing behavior in time, as discussed before. The distribution in the inner region, where the FRW metric is  valid, flattens out with time meaning that energy density is concentrating at the star center. At the final singularity ($t=t_0$), the mass function is constant everywhere.


\section{Beyond Spherical Symmetry}
\label{sec:genr spc}

In the following, we will explore how our AK-current approach can be used to construct conserved quantities in generic spacetimes .
We  begin by considering  the case of rigidly rotating
neutron stars, which allows to compare the resulting definition of
mass obtained through either the current $\,J^a_{T}\,$ or the almost-Killing current
$J^a_{AK}$. Finally, we will consider here
the generic case, in which the problem to find out a suitable AKV for
the current $J^a_{AK}$ is analyzed as a Cauchy problem.


\subsection{Rigidly Rotating Stars}
The spacetime of stationary and axisymmetric (uniformly rotating) stars can be obtained by solving
the hydro-Einstein equations  through a multi-domain
spectral code, as described in~\cite{Ansorg01,Ansorg:2003br}.  The spacetime may contain or not an ergoregion
depending on the compactness of the star.
The corresponding Lewis-Papapetrou coordinates are uniquely determined by the matching conditions at the surface (see ~\cite{Ansorg01,AnsorgBook08} for a detailed description). In the interior region we can write
\begin{eqnarray}
ds^2 &=&- e^{2\,U}\,(dt+a\,d\phi)^2 +e^{-2\,U} \left[ e^{2\,k}\,(d\rho^2+d\xi^2)
+W^2\,d\phi\,^2\right]\,.
\label{LP_interior}
\end{eqnarray}
where we have used  a  comoving coordinates system.

In this comoving frame, the stress-energy tensor is given by
\begin{equation}
T^{ab}=(\,\mu
+p\,)\,u^a\,u^b+p\,g^{ab}\,,\qquad
u^a= e^{-\,U}\,{\delta^a}_t\,,
\label{Tab_star}
\end{equation}
where $\mu= \rho\,(1+\epsilon)$ is the energy density and $p$ is the
pressure. It turns out that, given a particular equation of state, the
conservation of the stress-energy tensor~(\ref{Tab_star}) yields
\begin{equation}
e^{U}\,\text{exp}\,\left[\int_0^{p} \frac{dp}{\mu+p}
\right]=e^{V_0}=\text{const}\,.
\end{equation}
The compactness of the star is therefore  controlled through the parameter
$V_0$. We  have constructed several solutions  with an equation of state
for homogeneous matter with constant density $\mu=\text{const}$.
For all the stars the value of the spin parameter is $a=J/M^2 \approx 0.9$.
The mass, rotation frequency $\Omega$ and other parameters of the solutions
can be found  in table I of Ref.~\cite{Ruiz:2012te}.

Here again, since the resulting spacetime is axisymmetric  and stationary,
we can define the conserved currents  $J^a_{T}$ and
$J^a_{AK}$ as before. Note however that the choice of the seed KV is not
unique, as any linear combination of the two KVs is indeed a KV. The resulting
conserved quantities would depend then on the selected combination. We
will choose here the KV which is aligned with the fluid worldlines,
so that in comoving coordinates, it can be expressed as
$\,\xi^a = e^Uu^a = \delta^a_{(t)}$. Therefore, according
to~(\ref{Killing_current}) and (\ref{AKV3}), the conserved
currents are
\begin{eqnarray}
J_{T}^a&=&{T^a}_b\,\xi^b=-\mu\,e^U\,u^a\,,\\
J^a_{AK}&=&\frac{1}{4\pi}\,{R^a}_b\,\xi^b
=-\,\left(\mu+3\,p\,\right)\,e^U\,u^a\,.
\label{K_currents}
\end{eqnarray}
The associated energy densities are therefore
\begin{eqnarray}
\mathcal{E}_{T}&=&-n_a J_{T}^a=\mu\,A\,e^U\,,\\
\mathcal{E}_{AK}&=&-n_a J^a_{AK}
=\left(\mu+3\,p\,\right)A\,e^U\,,
\label{K_densities}
\end{eqnarray}
where $A=n\cdot u~$ is the Lorentz factor.

\begin{figure}[h]
\begin{center}
\includegraphics[width=80mm]{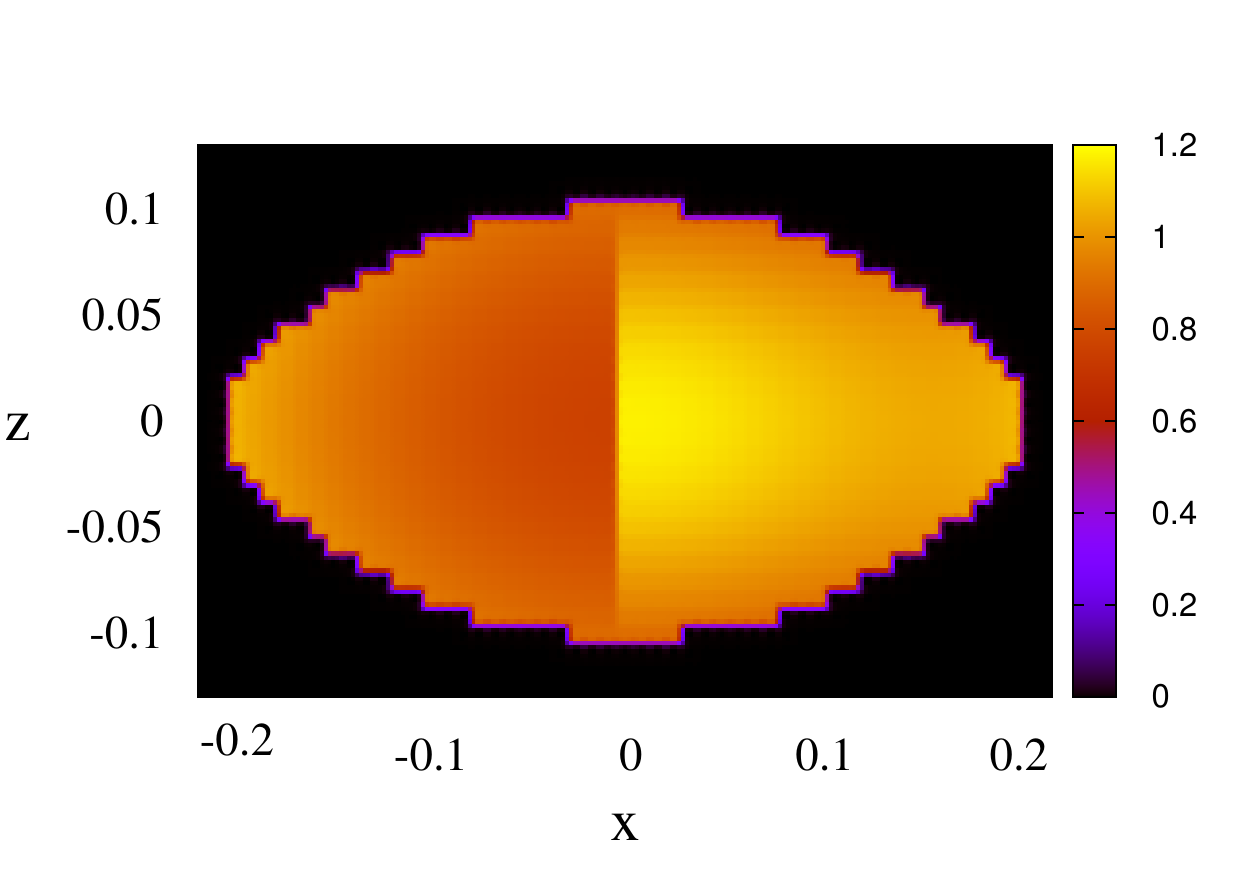}
\includegraphics[width=80mm]{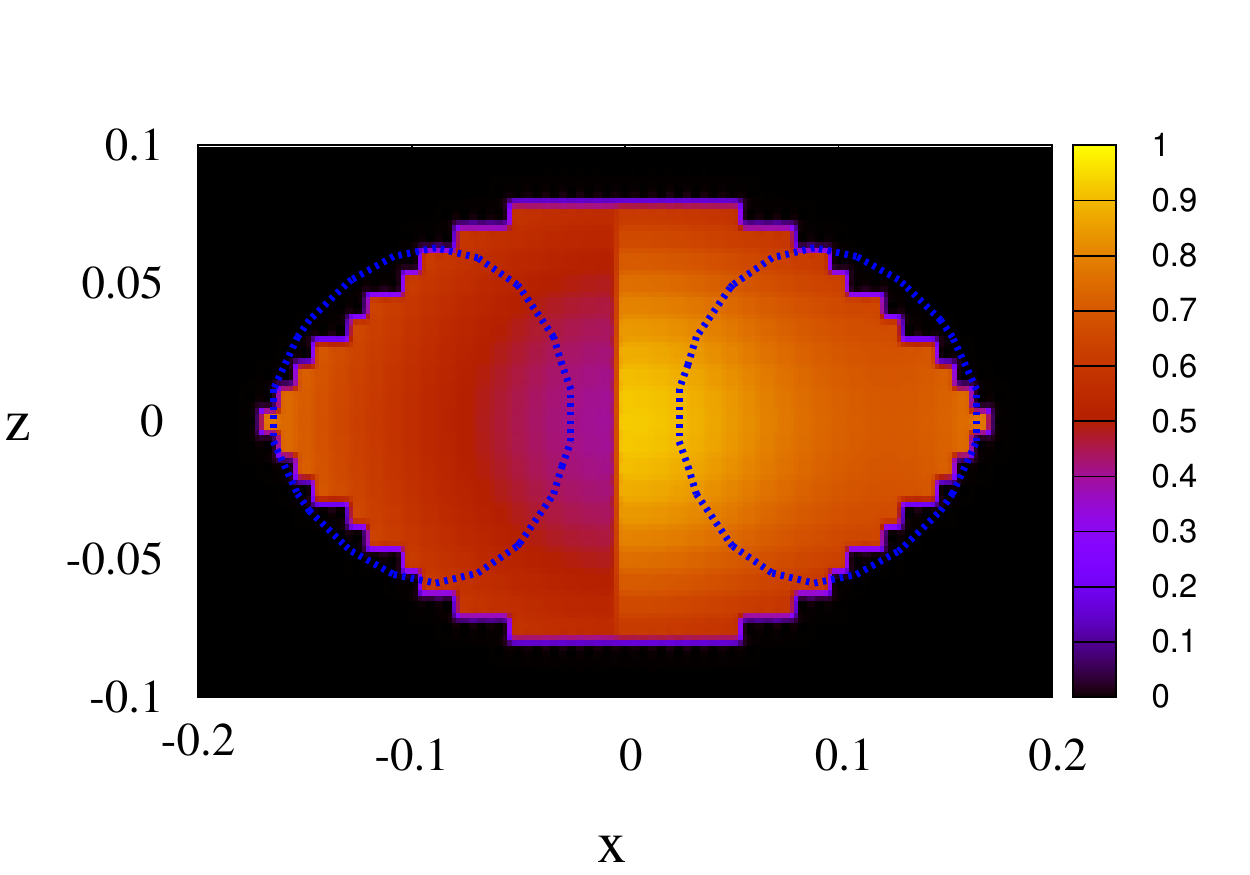}
\caption{Comparison between the energy densities $\mathcal{E}_{T}$
and  $\mathcal{E}_{AK}$ for a rapidly rotating neutron
stars with and without ergosphere (right and left panel, respectively) on the
vertical plane. The  dotted line
corresponds to the ergosphere surface, which has a torus-like shape.
In both configuration, the energy density $\mathcal{E}_{T}$ is displayed on the
negative $x$ axis, while that  AK energy density is plotted on the
positive side. The $\mathcal{E}_{AK}$ distribution
has a maximum at the center of the star that monotonically
decreases until the surface of the star while the $\mathcal{E}_{T}$ distribution
corresponds to a increasing function that reaches its maximum value at the
surface of the star.
}
\label{fig:Energy_DensAnsorg}
\end{center}
\end{figure}

In figure~\ref{fig:Energy_DensAnsorg} we compare the resulting mass-energy
densities on a vertical plane for the two cases $V_0=-0.8$ and $V_0=-1.25$
(see table I of Ref.~\cite{Ruiz:2012te}). The configuration displayed in the left
panel corresponds to a standard neutron star with constant density while the
configuration displayed  in the right panel corresponds a neutron star with
a torus-shaped ergoregion (signaled by a dotted line). 
The $\,\mathcal{E}_{T}\,$ energy density is displayed on the negative $x$ axis, while that
$\mathcal{E}_{AK}$ energy density is plotted on the positive side.
Notice that in both star models (with and without ergosphere) the $\,\mathcal{E}_{T}\,$
energy density has lower values close to the axis that increases
towards the surface of the star whereas the AK energy density has a maximum
at the center that decreases monotonically outward until the surface. This is the same behavior 
as the one obtained in section~\ref{sec:spherical case} for the spherical
case. The conclusion is again that the energy density $\mathcal{E}_{AK}$
matches the physically expected profile whereas $\,\mathcal{E}_{T}\,$ has a counter-intuitive
behavior.


\subsection{Generic Spacetimes}

As we have seen in section~\ref{sec:AKV}, the AKE~(\ref{AKE}) can always
be solved in a generic spacetime provided that $\mu \neq 2$.
Nevertheless, in this paper we are considering precisely the special case
$\mu=2$ because just in this case we obtain a suitable energy current $J_{AK}$.

Let us look now at the definition of the current~$J_{AK}$ which is
given in Eq.~(\ref{AK current detail}). It can be interpreted as the AKV differential equation for the $\mu=2$ case namely
\begin{equation}
\nabla_b\nabla^{[a}\xi^{b]}
= {R^a}_b\,\xi^b\,.
\label{AKV4}
\end{equation}
It is easy to show that the time derivative
of the time component of the vector field $\xi$
does not appear in the above system.
Only the antisymmetric combination of first derivatives is
involved. Therefore,
the principal symbol becomes singular in Fourier space~\cite{PhysRevD.72.124010}.
Although we have four second order equations for the four $\xi^a$ components, only
the space components $\xi^i$ of the AKV can be computed from their
evolution equations in a straightforward way.

This opens the door to consider the time component of (\ref{AKV2}) as a
constraint, in the same spirit as in the {\it free evolution approach} in
numerical relativity. Let us consider the quantities ${\cal C}^a$ representing
deviations from the Eq.~\ref{AKV4}:
\begin{equation}
\nabla_b\nabla^{[a}\xi^{b]} = {R^a}_b\,\xi^b + {\cal C}^a\,.
\end{equation}
If we take now the divergence of this vector relation, we obtain
\begin{equation}
\nabla_a{\cal C}^a = -\nabla_a({R^a}_b\,\xi^b) = -4\pi\,\nabla_a J_{AK}^a\,.
\end{equation}
Therefore, the conservation of the AK-current amounts to the conservation of the
deviations ${\cal C}^a$, that is
\begin{equation}\label{deviations}
\partial_0 (\sqrt{g}\, {\cal C}^0) + \partial_i (\sqrt{g}\, {\cal C}^i) = 0\,.
\end{equation}

In the vacuum case, where $J_{AK}$ vanishes, it would be enough to
impose the constraint ${\cal C}^0=0$ in the initial data in order to
ensure that it will hold during the whole evolution, provided that we
enforce ${\cal C}^i=0$ by computing the space components $\xi^i$ in
the right way, namely:
\begin{equation}\label{space AKV}
\nabla_b\nabla^{[b}\xi^{i]} + {R^i}_b\,\xi^b = 0\,.
\end{equation}
At the same time, one would get the some gauge freedom for choosing $\xi^0$.
Notice that in vacuum, the expression~(\ref{AKV4}) is equivalent to the first
set of Maxwell equations, in which the vector field $\xi$ would play the
role of the electromagnetic  potential. Nevertheless,
in the non-vacuum case the problem becomes circular since the conservation of
$J_{AK}$ is not granted unless $\xi$ is a true AKV and this requires indeed
${\cal C}^0=0$. There is no simple alternative to using the elliptic-type
equation
\begin{equation}\label{time AKV}
\nabla_a\nabla^{[a}\xi^{0]} + {R^0}_a\,\xi^a = 0\,,
\end{equation}
for computing the $\xi^0$ component. This (partially) elliptic nature of the
problem showed up yet in the spherical dust case (see section
\ref{sec:spherical case}), where we used the mass function
as a sort of potential for $\xi$, leading to the elliptic equation
(\ref{mass_eq}). In the generic (non-spherical) case, the system of AKV
equations is of a mixed elliptic-hyperbolic nature, like the ones considered
in recent Numerical Relativity developments~\cite{CorderoCarrion:2012ic}.
Studying the mathematical properties of this system is beyond the scope
of this work.


\section{Conclusions}
\label{sec:discussion}
We have proposed a new conserved current, describing energy densities of non-gravitational origin, namely
\begin{equation*}
J^a_{AK} = \frac{1}{4\pi}\,
{R^a}_b\,\xi^b\,.
\end{equation*}
We called it almost-Killing current because it requires (a particular case of) a timelike almost-killing vector $\xi$ in order to ensure conservation. The physical
meaning of such current is of course related with the physical meaning of the AKV itself.

As a first instance, we have explored the stationary/static case, in which
the  AKV can be chosen to be a Killing vector, which allows us to obtain
the well known current
\begin{equation*}
  J^a_{T} = {T^a}_b\,\xi^b\,.
\end{equation*}
Our results show that the energy density profiles obtained with $J_{AK}$
are the expected ones: the profile reaches a maximum value at the center of the
star and decreasing outwards until the surface of it. In strong contrast,
the energy associated with $J_T$ corresponds to an increasing
function which reaches its maximum value at the surface of the star.
We have shown these opposed behaviors for the Schwarzschild
(constant-density) star, as well as for rigidly rotating stationary stars
which can contain an ergoregion~\cite{Ansorg01}.
The unphysical behavior of $J_{T}$ is not a surprise. It is well known that, in order to recover the Hern\'andez-Misner mass function~\cite{Hernandez66}, the Killing vector must be rescaled in a non-trivial way (Kodama vector).
Our results confirm that the proposed current $J_{AK}$ provides
a possible solution for that problem in both the static and the stationary
cases.

Of course, there is an inherent ambiguity in our approach, as any AKV can be used as a seed for generating the corresponding conserved current. The stationary case
provides a good reference in this sense, because the ambiguity in the choice of
the seed AKV is solved by selecting precisely the timelike KV. In the stationary
axisymmetric case, however, some ambiguity reappears since the choice of the
KV is not unique. This ambiguity problem grows in the generic non-stationary case, which is
a bigger challenge in many respects.
Note however that in the spherical case, where a mass function $M_{AK}$ can be explicitly computed, one gets a simple generalization of the Kepler law (\ref{Mass Kepler}), strongly supporting the physical interpretation of our results, and this is so for any selection of the seed AKV.

We have considered the whole class of
Tolman-Bondi solutions for spherical balls of dust~\cite{KraHer80}. In
comoving coordinates, the Hern\'andez-Misner mass function for this
case is time-independent, $m=m(r)$, suggesting a baryonic mass
interpretation, although it is different from the standard result for dust, obtained from the matter current
\begin{equation*}
  J^a_{dust} = \rho\,u^a\,.
\end{equation*}

The mass function $M(r,t)$ associated to the proposed current
$J_{AK}$ depends instead on time in the generic case, which is the
expected behavior for the energy density in a dynamical collapse
scenario, where the kinetic energy of particles is varying in time. The static character of the Hern\'andez-Misner mass suggests that it is linked to the comoving observers, which are not quasi-stationary in the generic case.
We have shown that the proposed mass function can be used as a
potential, so that the AKV equations can be expressed as a single
elliptic-type equation for $M(r,t)$.

In the generic, non-spherical,
case the AKV set of equations is of a mixed elliptic-hyperbolic type,
like the ones considered in recent Numerical Relativity
developments~\cite{CorderoCarrion:2012ic}. We are currently working
on the properties of this system with a view to devising suitable
coordinate systems, adapted to the type of conservation
laws considered in the paper.


\acknowledgments This work was supported by Spanish Ministry of Science and
Innovation under grants CSD2007-00042, CSD2009-00064 and FPA2010-16495, and
Govern de les Illes Balears.

\bibliographystyle{apsrev}              
\bibliography{refs/references}{}        

\begin{thebibliography}{19}
\expandafter\ifx\csname natexlab\endcsname\relax\def\natexlab#1{#1}\fi
\expandafter\ifx\csname bibnamefont\endcsname\relax
  \def\bibnamefont#1{#1}\fi
\expandafter\ifx\csname bibfnamefont\endcsname\relax
  \def\bibfnamefont#1{#1}\fi
\expandafter\ifx\csname citenamefont\endcsname\relax
  \def\citenamefont#1{#1}\fi
\expandafter\ifx\csname url\endcsname\relax
  \def\url#1{\texttt{#1}}\fi
\expandafter\ifx\csname urlprefix\endcsname\relax\def\urlprefix{URL }\fi
\providecommand{\bibinfo}[2]{#2}
\providecommand{\eprint}[2][]{\url{#2}}

\bibitem[{\citenamefont{Ruiz et~al.}(2012)\citenamefont{Ruiz, Palenzuela,
  Galeazzi, and Bona}}]{Ruiz:2012te}
\bibinfo{author}{\bibfnamefont{M.}~\bibnamefont{Ruiz}},
  \bibinfo{author}{\bibfnamefont{C.}~\bibnamefont{Palenzuela}},
  \bibinfo{author}{\bibfnamefont{F.}~\bibnamefont{Galeazzi}}, \bibnamefont{and}
  \bibinfo{author}{\bibfnamefont{C.}~\bibnamefont{Bona}},
  \bibinfo{journal}{Mon.Not.Roy.Astron.Soc.} \textbf{\bibinfo{volume}{423}},
  \bibinfo{pages}{1300} (\bibinfo{year}{2012}), \eprint{1203.4125}.

\bibitem[{\citenamefont{Ansorg et~al.}(2002)\citenamefont{Ansorg, Kleinwachter,
  and Meinel}}]{Ansorg01}
\bibinfo{author}{\bibfnamefont{M.}~\bibnamefont{Ansorg}},
  \bibinfo{author}{\bibfnamefont{A.}~\bibnamefont{Kleinwachter}},
  \bibnamefont{and} \bibinfo{author}{\bibfnamefont{R.}~\bibnamefont{Meinel}},
  \bibinfo{journal}{Astron. Astrophys.} \textbf{\bibinfo{volume}{381}},
  \bibinfo{pages}{L49} (\bibinfo{year}{2002}), \eprint{astro-ph/0111080}.

\bibitem[{\citenamefont{Ansorg et~al.}(2003)\citenamefont{Ansorg, Kleinwachter,
  and Meinel}}]{Ansorg:2003br}
\bibinfo{author}{\bibfnamefont{M.}~\bibnamefont{Ansorg}},
  \bibinfo{author}{\bibfnamefont{A.}~\bibnamefont{Kleinwachter}},
  \bibnamefont{and} \bibinfo{author}{\bibfnamefont{R.}~\bibnamefont{Meinel}},
  \bibinfo{journal}{Astron.Astrophys.} \textbf{\bibinfo{volume}{405}},
  \bibinfo{pages}{711} (\bibinfo{year}{2003}), \eprint{astro-ph/0301173}.

\bibitem[{\citenamefont{Kodama}(1980)}]{Kodama:1979vn}
\bibinfo{author}{\bibfnamefont{H.}~\bibnamefont{Kodama}},
  \bibinfo{journal}{Prog. Theor. Phys.} \textbf{\bibinfo{volume}{63}},
  \bibinfo{pages}{1217} (\bibinfo{year}{1980}).

\bibitem[{\citenamefont{Hernandez~Jr and Misner}(1966)}]{Hernandez66}
\bibinfo{author}{\bibfnamefont{W.}~\bibnamefont{Hernandez~Jr}}
  \bibnamefont{and} \bibinfo{author}{\bibfnamefont{C.}~\bibnamefont{Misner}},
  \bibinfo{journal}{Astrophys. J.} \textbf{\bibinfo{volume}{143}},
  \bibinfo{pages}{452} (\bibinfo{year}{1966}).

\bibitem[{\citenamefont{Abreu and Visser}(2010)}]{Abreu:2010ru}
\bibinfo{author}{\bibfnamefont{G.}~\bibnamefont{Abreu}} \bibnamefont{and}
  \bibinfo{author}{\bibfnamefont{M.}~\bibnamefont{Visser}},
  \bibinfo{journal}{Phys. Rev.} \textbf{\bibinfo{volume}{D82}},
  \bibinfo{pages}{084023} (\bibinfo{year}{2010}), \eprint{gr-qc/10041456}.

\bibitem[{\citenamefont{Matzner}(1968)}]{Matzner:1968}
\bibinfo{author}{\bibfnamefont{R.~A.} \bibnamefont{Matzner}},
  \bibinfo{journal}{J. Math. Phys.} \textbf{\bibinfo{volume}{9}},
  \bibinfo{pages}{1657} (\bibinfo{year}{1968}).

\bibitem[{\citenamefont{Isaacson}(1968)}]{PhysRev.166.1263}
\bibinfo{author}{\bibfnamefont{R.~A.} \bibnamefont{Isaacson}},
  \bibinfo{journal}{Phys. Rev.} \textbf{\bibinfo{volume}{166}},
  \bibinfo{pages}{1263} (\bibinfo{year}{1968}),
  \urlprefix\url{http://link.aps.org/doi/10.1103/PhysRev.166.1263}.

\bibitem[{\citenamefont{Zalaletdinov}(2000)}]{Zalaletdinov99}
\bibinfo{author}{\bibfnamefont{R.}~\bibnamefont{Zalaletdinov}}
  (\bibinfo{publisher}{World Scientific}, \bibinfo{address}{Singapore},
  \bibinfo{year}{2000}).

\bibitem[{\citenamefont{Bona et~al.}(2005)\citenamefont{Bona, Carot, and
  Palenzuela-Luque}}]{PhysRevD.72.124010}
\bibinfo{author}{\bibfnamefont{C.}~\bibnamefont{Bona}},
  \bibinfo{author}{\bibfnamefont{J.}~\bibnamefont{Carot}}, \bibnamefont{and}
  \bibinfo{author}{\bibfnamefont{C.}~\bibnamefont{Palenzuela-Luque}},
  \bibinfo{journal}{Phys. Rev. D} \textbf{\bibinfo{volume}{72}},
  \bibinfo{pages}{124010} (\bibinfo{year}{2005}),
  \urlprefix\url{http://link.aps.org/doi/10.1103/PhysRevD.72.124010}.

\bibitem[{\citenamefont{Taubes}(1978)}]{Taubes78}
\bibinfo{author}{\bibfnamefont{C.~H.} \bibnamefont{Taubes}},
  \bibinfo{journal}{J. Math. Phys.} \textbf{\bibinfo{volume}{19}}
  (\bibinfo{year}{1978}).

\bibitem[{\citenamefont{York}(1974)}]{York74}
\bibinfo{author}{\bibfnamefont{J.~W.} \bibnamefont{York}},
  \bibinfo{journal}{Ann. Inst. H. Poincar\'e Sect. A}
  \textbf{\bibinfo{volume}{21}}, \bibinfo{pages}{319} (\bibinfo{year}{1974}).

\bibitem[{\citenamefont{Smarr and York}(1978)}]{PhysRevD.17.2529}
\bibinfo{author}{\bibfnamefont{L.}~\bibnamefont{Smarr}} \bibnamefont{and}
  \bibinfo{author}{\bibfnamefont{J.~W.} \bibnamefont{York}},
  \bibinfo{journal}{Phys. Rev. D} \textbf{\bibinfo{volume}{17}},
  \bibinfo{pages}{2529} (\bibinfo{year}{1978}),
  \urlprefix\url{http://link.aps.org/doi/10.1103/PhysRevD.17.2529}.

\bibitem[{\citenamefont{Komar}(1959)}]{PhysRev.113.934}
\bibinfo{author}{\bibfnamefont{A.}~\bibnamefont{Komar}},
  \bibinfo{journal}{Phys. Rev.} \textbf{\bibinfo{volume}{113}},
  \bibinfo{pages}{934} (\bibinfo{year}{1959}),
  \urlprefix\url{http://link.aps.org/doi/10.1103/PhysRev.113.934}.

\bibitem[{\citenamefont{Wald}(1984)}]{Wald84}
\bibinfo{author}{\bibfnamefont{R.~M.} \bibnamefont{Wald}},
  \emph{\bibinfo{title}{General Relativity}} (\bibinfo{publisher}{The
  University of Chicago Press}, \bibinfo{address}{Chicago, U.S.A.},
  \bibinfo{year}{1984}).

\bibitem[{\citenamefont{Bondi}(1947)}]{Bon52}
\bibinfo{author}{\bibfnamefont{H.}~\bibnamefont{Bondi}},
  \bibinfo{journal}{Monthly Notices of the Royal Astronomical Society}
  \textbf{\bibinfo{volume}{107}}, \bibinfo{pages}{410} (\bibinfo{year}{1947}).

\bibitem[{\citenamefont{D.~Kramer and Herlt}(1980)}]{KraHer80}
\bibinfo{author}{\bibfnamefont{M.~M.} \bibnamefont{D.~Kramer},
  \bibfnamefont{H.~Stephani}} \bibnamefont{and}
  \bibinfo{author}{\bibfnamefont{E.}~\bibnamefont{Herlt}},
  \emph{\bibinfo{title}{Exact Solutions of {E}instein's Field Equations}}
  (\bibinfo{publisher}{Cambridge University Press},
  \bibinfo{address}{Cambridge}, \bibinfo{year}{1980}).

\bibitem[{\citenamefont{{Meinel} et~al.}(2008)\citenamefont{{Meinel}, {Ansorg},
  {Kleinw{\"a}chter}, {Neugebauer}, and {Petroff}}}]{AnsorgBook08}
\bibinfo{author}{\bibfnamefont{R.}~\bibnamefont{{Meinel}}},
  \bibinfo{author}{\bibfnamefont{M.}~\bibnamefont{{Ansorg}}},
  \bibinfo{author}{\bibfnamefont{A.}~\bibnamefont{{Kleinw{\"a}chter}}},
  \bibinfo{author}{\bibfnamefont{G.}~\bibnamefont{{Neugebauer}}},
  \bibnamefont{and}
  \bibinfo{author}{\bibfnamefont{D.}~\bibnamefont{{Petroff}}},
  \emph{\bibinfo{title}{{Relativistic Figures of Equilibrium}}}
  (\bibinfo{year}{2008}).

\bibitem[{\citenamefont{Cordero-Carrion and
  Cerda-Duran}(2012)}]{CorderoCarrion:2012ic}
\bibinfo{author}{\bibfnamefont{I.}~\bibnamefont{Cordero-Carrion}}
  \bibnamefont{and}
  \bibinfo{author}{\bibfnamefont{P.}~\bibnamefont{Cerda-Duran}}
  (\bibinfo{year}{2012}), \eprint{1211.5930}.

\end{thebibliography}

\end{document}